\documentstyle[epsfig]{mn}

\newcommand{\gta}{\mathrel{\hbox{\rlap{\lower.55ex \hbox {$\sim$}}
                   \kern-.3em \raise.4ex \hbox{$>$}}}}
\newcommand{\lta}{\mathrel{\hbox{\rlap{\lower.55ex \hbox {$\sim$}}
                   \kern-.3em \raise.4ex \hbox{$<$}}}}

\newif\ifAMStwofonts

\ifoldfss
  \ifCUPmtlplainloaded \else
    \NewTextAlphabet{textbfit} {cmbxti10} {}
    \NewTextAlphabet{textbfss} {cmssbx10} {}
    \NewMathAlphabet{mathbfit} {cmbxti10} {} 
    \NewMathAlphabet{mathbfss} {cmssbx10} {} 
  \fi
  \ifAMStwofonts
    \ifCUPmtlplainloaded \else
      \NewSymbolFont{upmath} {eurm10}
      \NewSymbolFont{AMSa} {msam10}
      \NewMathSymbol{\upi}     {0}{upmath}{19}
      \NewMathSymbol{\umu}     {0}{upmath}{16}
      \NewMathSymbol{\upartial}{0}{upmath}{40}
      \NewMathSymbol{\leqslant}{3}{AMSa}{36}
      \NewMathSymbol{\geqslant}{3}{AMSa}{3E}

    \fi
  \fi
\fi 

\ifnfssone
  \newmathalphabet{\mathit}
  \addtoversion{normal}{\mathit}{cmr}{m}{it}
  \addtoversion{bold}{\mathit}{cmr}{bx}{it}
  \newmathalphabet{\mathbfit} 
  \addtoversion{normal}{\mathbfit}{cmr}{bx}{it}
  \addtoversion{bold}{\mathbfit}{cmr}{bx}{it}
  \newmathalphabet{\mathbfss} 
  \addtoversion{normal}{\mathbfss}{cmss}{bx}{n}
  \addtoversion{bold}{\mathbfss}{cmss}{bx}{n}
  \ifAMStwofonts
    \ifCUPmtlplainloaded \else
      \UseAMStwoboldmath
      \makeatletter
      \new@mathgroup\upmath@group
      \define@mathgroup\mv@normal\upmath@group{eur}{m}{n}
      \define@mathgroup\mv@bold\upmath@group{eur}{b}{n}
      \edef\UPM{\hexnumber\upmath@group}
      \new@mathgroup\amsa@group
      \define@mathgroup\mv@normal\amsa@group{msa}{m}{n}
      \define@mathgroup\mv@bold\amsa@group{msa}{m}{n}
      \edef\AMSa{\hexnumber\amsa@group}
      \makeatother
      \mathchardef\upi="0\UPM19
      \mathchardef\umu="0\UPM16
      \mathchardef\upartial="0\UPM40
      \mathchardef\leqslant="3\AMSa36
      \mathchardef\geqslant="3\AMSa3E
    \fi
  \fi
\fi 
\ifnfsstwo
  \DeclareMathAlphabet{\mathbfit}{OT1}{cmr}{bx}{it}
  \SetMathAlphabet\mathbfit{bold}{OT1}{cmr}{bx}{it}
  \DeclareMathAlphabet{\mathbfss}{OT1}{cmss}{bx}{n}
  \SetMathAlphabet\mathbfss{bold}{OT1}{cmss}{bx}{n}
  \ifAMStwofonts
    \ifCUPmtlplainloaded \else
      \DeclareSymbolFont{UPM}{U}{eur}{m}{n}
      \SetSymbolFont{UPM}{bold}{U}{eur}{b}{n}
      \DeclareSymbolFont{AMSa}{U}{msa}{m}{n}
      \DeclareMathSymbol{\upi}{0}{UPM}{"19}
      \DeclareMathSymbol{\umu}{0}{UPM}{"16}
      \DeclareMathSymbol{\upartial}{0}{UPM}{"40}
      \DeclareMathSymbol{\leqslant}{3}{AMSa}{"36}
      \DeclareMathSymbol{\geqslant}{3}{AMSa}{"3E}
    \fi
  \fi
\fi 

\ifCUPmtlplainloaded \else
  \ifAMStwofonts \else 
    \def\upi{\pi}
    \def\umu{\mu}
    \def\upartial{\partial}
  \fi
\fi

\newbox\grsign \setbox\grsign=\hbox{$>$}
\newdimen\grdimen \grdimen=\ht\grsign
\newbox\laxbox \newbox\gaxbox
\setbox\gaxbox=\hbox{\raise.5ex\hbox{$>$}\llap
     {\lower.5ex\hbox{$\sim$}}}\ht1=\grdimen\dp1=0pt
\setbox\laxbox=\hbox{\raise.5ex\hbox{$<$}\llap
     {\lower.5ex\hbox{$\sim$}}}\ht2=\grdimen\dp2=0pt
\def\gax{\mathrel{\copy\gaxbox}}
\def\lax{\mathrel{\copy\laxbox}}
\def\lta{\lax}
\def\gta{\gax}

\def\WZ{{WZ\,Sge\ }}

\title[WZ\,Sge]
  {WZ Sagittae as a DQ Herculis star}

\author[Lasota, Kuulkers \&\ Charles]
  {Jean-Pierre~Lasota$^{1}$\thanks{E-mail:
  jpl@orge.obspm.fr (JPL), E.Kuulkers@sron.nl (EK), pac@astro.ox.ac.uk (PC).}\thanks{Present address:
  Institut d'Astrophysique de Paris, CNRS, 98bis Boulevard Arago, 75014 Paris, France},
  Erik Kuulkers$^{2\star}$\thanks{Present address: Space Research 
  Organization Netherlands (SRON), Sorbonnelaan 2, 3584 CA Utrecht, 
  The Netherlands} 
  \&\ Phil Charles$^{2\star}$\\
  $^1$UPR, 176 du CNRS, DARC, Observatoire de Paris, Section de Meudon,
  F-92195 Meudon Cedex, France\\
  $^2$Astrophysics, University of Oxford, Nuclear and Astrophysics Laboratory, 
  Keble Road, Oxford OX1 3RH, United Kingdom}

\pagerange{\pageref{firstpage}--\pageref{lastpage}}
\pubyear{1998}

\begin{document}

\maketitle

\label{firstpage}

\begin{abstract}
We argue that quiescent WZ Sge is a rapidly spinning magnetic rotator
in which most of the matter transfered from the secondary is ejected
from the system. Assuming that the observed 27.87 s oscillation period
is due to the spinning white dwarf we propose that the other observed
principal period of 28.96 s is a beat due to reprocessing of the
rotating white dwarf beam on plasma blobs in Keplerian rotation at the
outer disc rim. The weaker, transient, 29.69 s period is identified as
a beat with the Keplerian period of the magnetosphere. WZ Sge evolves
through a cycle of spin-up and spin-down phases. During the spin-down
phase it is a DQ Her star, during the spin-up phase it should be a ER
UMa star.
\end{abstract}


\section{Introduction}

WZ Sagittae is a remarkable dwarf nova binary system.  Dwarf novae are
a subclass of cataclysmic variables (CVs) which show more or less
regular outbursts. In CVs, a late-type, Roche-lobe filling secondary
star loses mass which, in general, is accreted by a white dwarf
primary. Usually, when the white dwarf is not too strongly magnetized
the accreting matter forms an accretion disc.  It is well established
that such discs are the sites of dwarf nova outbursts. \WZ outbursts
are very rare (the recurrence time is around 30 years) and they are
only of the `superoutburst' type, i.e. they are long ($\sim$ 30 days),
high amplitude ($\sim$ 7 mag) outbursts during which a `superhump' in
the optical light-curve is observed at a period slightly longer than the
orbital one. \WZ is the prototype of a class of systems with similar
outburst properties (sometimes also called `TOADs'). The orbital period
of WZ Sge is 81.63 minutes, close to the minimum period ($\sim 80$ min)
below which no (non-degenerate) CV has been observed.

It is generally believed  that dwarf nova outbursts result from a
thermal-viscous instability present in accretion discs at temperatures
corresponding to hydrogen partial ionization. In the standard version
of the disc instability model (DIM; see Hameury et al. 1998 for the
most recent version of this model, Cannizzo 1993 and Lasota \& Hameury
1998 for  reviews) one assumes that  mass-transfer from the secondary
is constant prior to, and during the outburst.  Properties of outbursts
depend on the viscosity mechanism which transports angular momentum and
heats the disc.  In the DIM the kinematic viscosity coefficient is
taken as $\nu=\alpha c_s H$, where $c_s$ is the (adiabatic) speed of
sound, $H$ the disc semi-thickness, and the `viscosity coefficient'
$\alpha < 1$ (Shakura \& Sunyaev 1973).  The DIM requires different
values of $\alpha$'s in outburst and in quiescence (Smak 1984b). The
quiescent value for the great majority of systems should be $\sim 0.01$
(Livio \& Spruit 1991). Smak (1993) showed, however, that the outburst
cycle of WZ Sge cannot be described by such a version of the DIM (see
also Osaki 1996). The main reason is that for $\alpha \sim 0.01$ the
disc is able to accumulate during quiescence only $\sim 10^{21}$ g
whereas during outbursts $\sim 10^{24}$ g is accreted by the white
dwarf. Assuming that $\alpha \sim 10^{-5}$ solves this inconsistency
and the resulting very low viscosity may also account for the long
recurrence time which is difficult to reproduce with the `standard'
$\alpha$ values. The difficulty of this version of the model is that
one must find an explanation for the very low value of the viscosity
coefficient in this particular system. The low value of the
mass-transfer rate ($\sim 10^{15}$ g s$^{-1}$, invoked as an
explanation; Osaki 1996) is not much lower than in other systems with
similar orbital periods so it is doubtful that it can explain a
difference in $\alpha$ by three orders of magnitude. In any case the
relation between accretion rate and viscosity, if any, is unknown (see
Gammie \& Menou 1998).

A different modification of the standard DIM was suggested by Lasota,
Hameury \& Hur\'e (1995) and Hameury, Lasota \& Hur\'e (1997).  They
assume that the value of the $\alpha$-parameter in WZ Sge is not
different from the `standard' one. They noticed that at low mass
transfer rates outer regions of accretion discs in CVs are cold enough
to be stable with respect to the thermal instability believed to be
responsible for dwarf-nova outbursts.  They proposed that the WZ Sge
accretion  disc  does not extend down to the white dwarf's surface but
is truncated at a radius corresponding to a stable outer disc.  An
enhancement of mass transfer would bring such a disc into an unstable
state and thus trigger an outburst. One should stress that such an
outburst would still be due to the thermal-viscous instability, the
enhanced mass transfer serves only as a trigger.  The recurrence time
is then related to the characteristic time-scale of mass-transfer
fluctuations. One can also obtain long recurrence times if the
truncated disc is marginally unstable (Warner, Livio \& Tout  1996) but
in practice such a disc is indistinguishable from a marginally stable
one proposed by Lasota et al. (1995). There is still, however, a problem
to be solved: since $\alpha \sim 0.01$ there would be only $\sim
10^{21}$ g in the disc available. Therefore, in this case mass must be
added to the disc {\sl during} outburst.  There is observational
evidence (Smak 1997; see also Hessman et al. 1984) that irradiation of
the secondary during outburst increases mass transfer from the
secondary (see Hameury et al. 1997 for a simple model).

In Lasota et al. (1995) and Hameury et al. (1997) the truncation was
assumed to be due either to evaporation (Meyer \& Meyer-Hofmeister
1994) or to the presence of a magnetic field
strong enough to disrupt the quiescent disc ($\gta 10^4$ G would be
sufficient for a low mass white dwarf). The second hypothesis seemed to
be favoured by the presence, before the last, 1978 outburst, of a 27.87
s coherent oscillation which could be attributed to the rotation of an
accreting white dwarf (Patterson 1980, hereafter P80).  There were two
problems with this interpretation.  First, in addition to the 27.87 s
period several other (some of them transient) `satellite period'
oscillations have been observed (Robinson, Nather \& Patterson 1978;
P80). The interpretation of these periods in terms of beat frequencies
resulting from reprocessing of the white dwarf's pulsed light on various
features in the accretion flow encountered several difficulties (P80) .
Second, all the oscillations disappeared during the outburst.
After the outburst the principal, 27.87 s, oscillation has been absent
for 16 years (although the 28.96 s was seen when \WZ was at twice its
pre-outburst brightness). Although it was easy to understand that during
the outburst the increased accretion rate could suppress the
magnetosphere, the lack of the principal pulse afterwards was difficult
to understand and cast a shadow of doubt on the presence of a rapidly
rotating white dwarf in WZ Sge.

In 1995 the 27.87 s oscillation reappeared in the company of the
previously present 28.96~s pulsation (Patterson et al. 1998, hereafter
P98).  Weak, transient `satellite period' pulses are also seen at 28.2
and 29.69 s.  The most important news, however, was the detection of a
$27.86\pm 0.01$ s period in the 2-6 keV ASCA energy band (P98) which
confirmed the presence of a rotating, magnetized white dwarf in WZ
Sge.  The presence of a rapidly rotating white dwarf was confirmed by
UV spectral observations (Cheng et al. 1997). \WZ is therefore a DQ Her
star, a CV containing a rapidly rotating magnetized white dwarf.

In a recent article Meyer-Hofmeister, Meyer \& Liu (1998) (see also
Mineshige et al. 1998) argue that the inner disc `hole' in WZ Sge is
due to evaporation and not to the presence of a magnetic field. These
authors, however, underestimate by a significant factor the coherence
of the observed pulsations and their model seems to be contradicted by
observations. For example, their model predicts the presence of
superhumps well before the beginning of the superoutburst whereas in WZ
Sge this feature appeared only 10 - 12 day {\sl after} the maximum
\cite{p81}. Also the remarkable constancy of the X-ray luminosity 
after the 1978 outburst is in contradiction with their model.

In this article we argue that Patterson's (P80, also P98) hypothesis
that WZ Sge contains a rapid, oblique magnetic rotator is consistent
with most of the observed properties of this system. We show that
sideband oscillations observed in \WZ can be interpreted as resulting
from reprocessing of the white dwarf's pulsed light on various features
of the transfered flow. In Section 2 we present the observed properties
of \WZ and identify various oscillations with orbital sidebands. In
Section 3 we discuss  possible configurations of WZ Sge as a system
containing a magnetic rotator and  show that only one is consistent
with the system properties at outburst. In Section 4 we discuss the
accretion history of WZ Sge and its relation to other CVs.  We
summarize our model in Section 5. 

\section{\WZ\ as a magnetic rotator}

If \WZ is an oblique magnetic rotator one should, in principle, observe
in its optical light-curve the fundamental spin frequency and some
sidebands resulting from the beat between this and the orbital
frequencies (Warner 1986). In what follows we will assume that all
characteristic periods observed in WZ Sge have a rotational origin. The
possibility that these oscillations (or some of them, Warner 1995a) are
due to white dwarf pulsations cannot be excluded
(see Wood 1999 for recent arguments in favour of this model). Such a model is,
however, hard to test because it does not make any predictions (P98).
In what follows we will try to show that all observed periodicities
observed in \WZ can be explained in the framework of a magnetic rotator
model.

\subsection{General properties and parameters}

Following Warner (1995a) we classify \WZ as a DQ Her star. DQ Her stars
are rapidly rotating intermediate polars (IPs), i.e. magnetic CVs in
which the white dwarf rotation is not synchronous with the orbital
motion (synchronous systems are called `polars' or `AM Her stars').
Warner (1995a) adds also that DQ Her stars are characterized by the
absence of hard X-rays in the sense that, contrary to `usual' IPs their
X-ray temperature is much lower than 10 keV. In \WZ the X-ray
temperature is $\sim 4.5$ keV (P98), so it is a hard X-ray emitter,
even if its temperature is lower than that in `usual' IPs (see below).

\WZ has been one of the best observed CVs but the values of its
fundamental parameters are still uncertain. Smak's (1993) photometric
solution gives $M_1 = 0.45$ (where $M_1$ is the white dwarf mass in
solar units) and $q \equiv M_2/M_1= 0.13$, where $M_2$ is the
secondary's mass, whereas Spruit \& Rutten (1998), who model the `hot
spot' at which the mass transfer stream interacts with the outer disc
regions, get $M_1 = 1.2$ and $q=0.075$. As we shall see Smak's value is
too small if the white dwarf in \WZ is spinning at 27.87 s, but the
Spruit \& Rutten (1998) value might be too high.

In Figure 1 we show various mass-radius relations relevant for WZ Sge.
The white dwarf mass-radius relation is that of Nauenberg (1972) which
is suitable for helium white dwarfs (using other mass-radius relations
gives very similar results, e.g. P98). Also plotted is the corotation radius
\begin{equation}
R_{\Omega} = \left({GM_{\odot}M_1 P^2_s \over 4 \pi^2}\right)^{1/3}=
             1.5 \times 10^8 P_s^{2/3} M_1^{1/3} {\rm cm},
\label{corot}
\end{equation}
where $P_s$ is the white dwarf spin period. The corotation radius
corresponds to a distance at which a free particle in circular Keplerian
orbit corotates with the
white dwarf.  In particular the white dwarf radius must satisfy $R_1 <
R_{\Omega}$.  If one assumes that $P_s = 27.87$ s, Figure 1 shows that
for Smak's $M_1=0.45$ the white dwarf would be rotating just below the
break-up speed so $M_1$ must be larger than 0.45 (see also P98).

\begin{figure} 
\centering\epsfig{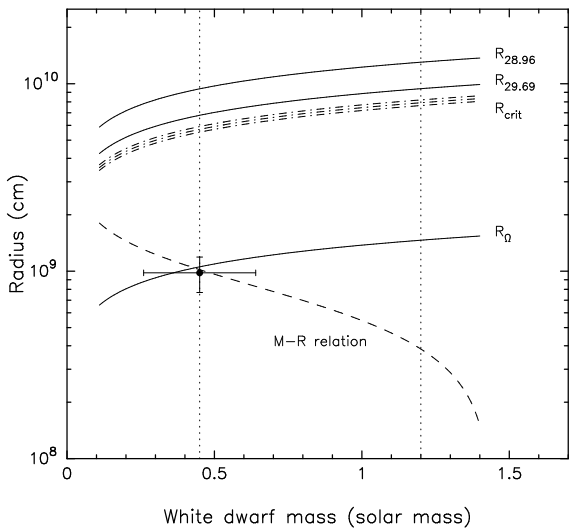} 
\caption{
Various radius vs.\ mass relations appropriate for WZ Sge. Shown are the
mass-radius (M-R) relation for helium white dwarfs 
(Nauenberg 1972; dashed line), the corotation radius ($R_{\rm \Omega}$; Eq.~1),
the critical radius ($R_{\rm crit}$) above which a disc is marginally stable
(Eq.~6) for 3 values of $\alpha$ (see text; dash-dotted lines), and
the Kepler radii at which the beat between the white dwarf spin 
frequency and the Kepler frequency gives rise to the observed oscillations at 
28.96\,s and 29.69\,s. The inferred solution from Smak (1993) is also 
indicated. With dotted lines we indicate the white dwarf masses as derived by
Smak (1993) and Spruit \&\ Rutten (1998), i.e.\ 0.45\,M$_{\odot}$ and
1.2\,M$_{\odot}$, respectively.
} 
\label{1} 
\end{figure} 

\subsection{Orbital sidebands}

In addition to the 27.87 s oscillation several other optical
oscillations between 27.8 and 30\,s have been observed in WZ\,Sge (P98,
and references therein).  In general either the 27.87 or 28.96\,s
oscillations are present, sometimes with additional periods up to
$\sim$30\,s (i.e.\ 28.19\footnote{The 28.19\,s oscillation might be
composed of signals at 28.14\,s and 28.24\,s (P80).}, 28.52,
29.69\,s). A 28.2\,s oscillation is also seen in the UV (Welsh et
al.\ 1997) (strictly speaking the UV oscillation is $28.09\pm 0.12$,
but we will follow P98 in identifying it with the optical 28.2 s (28.19
s) period).

We will denote the white dwarf angular rotation frequency with
$\omega=2\pi/P_s$ and the orbital frequency as $\Omega=2\pi/P_{\rm
orb}$.  As shown by Warner (1986), in intermediate polars the rotating
white dwarf's beam and its reprocessing on the surface layers of the
disc, hot spot and/or the secondary produces peaks in the power spectra
of optical light curves not only at $\omega$ and $\omega$$-$$\Omega$,
but there also exist sidebands at $\omega$+$\Omega$ and
$\omega$$-$$2\Omega$. (Wynn \& King 1992 analyzed  power spectra of
X-ray light curves in IPs in the case of discless accretion.) For a
white dwarf rotation period of 27.87\,s and an orbital period of
81.63\,min (see P98) this gives sidebands at 28.03, 27.71 and
28.19\,s.  Patterson (1980) already noted that the 28.2\,s period is
consistent with $\omega$$-$$2\Omega$.  He also noted that significant
peaks in the power spectra exist in the sidelobes close to the main
peak at 27.87\,s. These periods were, however, not discussed by him;
visual inspection of his Figure~3b shows that the peaks of these
sidelobes are at $\sim$27.96 and $\sim$27.79\,s.  These are very close
to $\omega$$-$$\Omega$ and $\omega$+$\Omega$, respectively. We
therefore propose that the three optical sideband frequencies are
present, making stronger the suggestion that an appreciable magnetic
field exists in WZ\,Sge, truncating the inner parts of an accretion
disc and making it similar to an IP.  It also confirms that the
27.87\,s period is the white dwarf spin period; checks made by assuming
the white dwarf spin period is at the other observed frequencies show
that although the expected sideband frequencies match the observed
frequencies there is the no explanation for the 27.87\,s period. For
example, assuming 28.19 s as the white dwarf spin period gives
sidebands at 28.03, 28.35, 28.52 s. These are all observed in the power
spectra of P80, P98. In addition $\omega - 3\Omega$ gives 28.96 s. $P_s
= 28.09$ s gives sidebands at 28.25, 27.93, 28.42 s, which in principle
could match the observed peaks in the power spectrum, as well as
$\omega - 3\Omega$ which gives 28.58 s.  In both cases, however, the
27.87 s oscillation remains unexplained.

We identify the 28.2 s optical and UV oscillations with
$\omega$$-$$2\Omega$ which would result from reprocessing on the hot
spot. Welsh et al.\ (1997) identify this period with the white dwarf
spin period but the presence of the 27.87\,s period in X-rays does not
support this identification. If the 28.2 s oscillation results from
reprocessing on the hot spot one must then explain why the fundamental
spin period (27.87\,s) is not seen in the UV. In order to understand
this absence one should, presumably, understand the structure of the
accretion flow. As we shall see below the spin period of \WZ implies that
this system is in an `ejector' phase, ejecting most of the transferred
matter. Such a case would be totally different from the usual IPs (see
e.g. Hellier 1996). No model of such a flow seems to exist at present.

It is also interesting to note that a reprocessing projected area might
also vary with the 2$\Omega$ component (Warner 1995a).  This will
introduce components at frequencies $\omega$+$\Omega$ and
$\omega$$-$$3\Omega$ in the power spectrum. The latter produces a peak
near 28.35\,s, which is visible (at $\sim$ 28.34\,s) as a shoulder of
the peak near 28.19\,s in the 'grand average' power spectrum of the
observations prior to the outburst end of 1978 (P80; see also P98).  We
note that the observed oscillation at 28.52\,s is consistent with
$\omega$$-$$4\Omega$ (P80) but have no interpretation for this fact.

Having explained to our satisfaction various sidebands we are still left
with the 28.96\,s and the 29.69\,s periods which are not commensurable
with the orbital period. The 28.96\,s signal appears as one of the two
main peaks in the power spectrum and is generally present. The 29.69\,s
is weak and transitory.

\section{\WZ as a magnetic ejector}

\subsection{A discless system ?}

Patterson (1980) suggested that the 28.96\,s period could result from the beat
between the white dwarf spin frequency and the Keplerian frequency at
the magnetosphere.  The required Keplerian period is $P_{\rm K}=$
733.47\,s and could correspond to the rotation period of plasma blobs
at the disc inner edge. The problem with such an interpretation is
that, if true, not much of a disc would be left. The corresponding
magnetospheric radius would be
\begin{equation}
R_{M}(28.96) = 1.22 \times 10^{10} M_1^{1/3} {\rm cm}.
\label{mrad1}
\end{equation}
According to Smak (1993), for $M_1=0.45$, the outer disc radius is 
$R_{\rm D} = (1.07\pm 0.19) \times 10^{10} {\rm cm}$, 
whereas Spruit \& Rutten (1998) obtain, for $M_1=1.2$,
$R_{\rm D} = (1.75\pm 0.14) \times 10^{10} {\rm cm}$ so that in both cases we 
would rather have a ring then a disc, or no disc at all, depending on the
details of plasma interaction with the magnetic field (see e.g. King 1993).
\begin{figure} 
\centering\epsfig{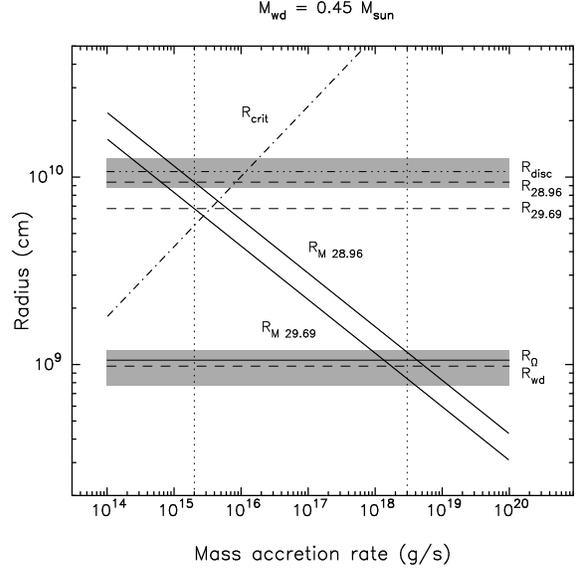} 
\caption{
Shown are the critical radius (Eq.~6, with $\alpha = 0.01$; dash-dotted
line) and the magnetospheric radius (Eq.~4) as a function of the mass
accretion rate for a white dwarf mass $M_{\rm wd}=0.45$\,$M_{\odot}$.
The magnetospheric radius relations are normalized so that the
magnetospheric radii in quiescence are equal to the Kepler radii which
are derived from the observed 28.96\,s and 29.69\,s pulsations. Also
indicated are the white dwarf radius ($R_{\rm wd}$; dashed line) and
outer disc radius ($R_{\rm disc}$; dash-dot-dotted line) as derived by
Smak (1993) and the corotation ($R_{\rm \Omega}$) and Kepler radii at which
the beat between the white dwarf spin frequency and the Kepler
frequency gives rise to the observed oscillations at 28.96\,s and
29.69\,s (dashed lines).  The uncertainties in $R_{\rm wd}$ and $R_{\rm
disc}$ are given by the grey hatched areas. The mass accretion rates of
WZ Sge during quiescence ($\sim$2\,10$^{15}$\,g\,s$^{-1}$) and during
the peak of its 1978 outburst ($\sim$3\,10$^{18}$\,g\,s$^{-1}$) as
estimated by Smak (1993) are indicated with dotted lines.
} 
\label{2a} 
\end{figure}
\begin{figure} 
\centering\epsfig{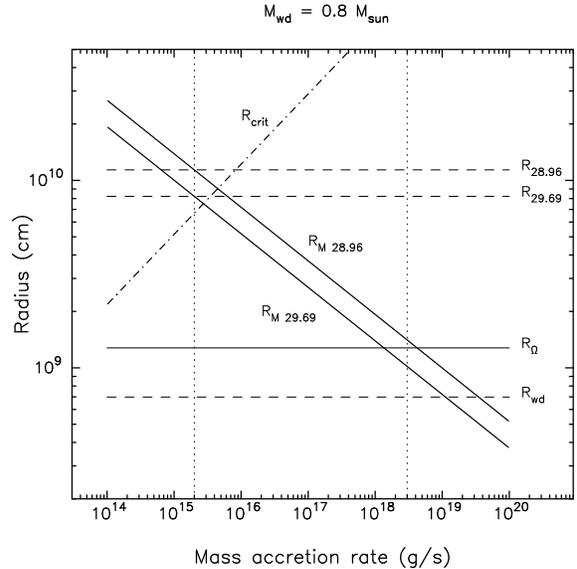} 
\caption{
As Figure 2, but now for M$_{\rm wd}=0.8$\,M$_{\odot}$. $R_{\rm wd}$ is 
determined from the mass-radius relation appropriate for helium
white dwarfs (Nauenberg 1972). No solution of $R_{\rm disc}$ exists in the
literature for M$_{\rm wd}=0.8$\,M$_{\odot}$.
} 
\label{2b} 
\end{figure}
\begin{figure} 
\centering\epsfig{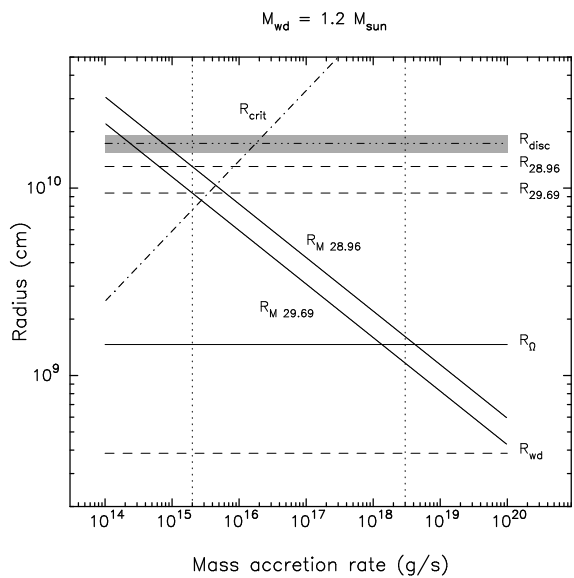} 
\caption{
As Figure 2, but now for M$_{\rm wd}=1.2$\,M$_{\odot}$. $R_{\rm wd}$ is 
determined from the mass-radius relation appropriate for helium
white dwarfs (Nauenberg 1972). $R_{\rm disc}$ (dash-dot-dotted line) is 
the solution given by Spruit \&\ Rutten (1998); its uncertainty is indicated
by the grey hatched area.
} 
\label{2c} 
\end{figure}
In addition, if the Keplerian radius $R_{\rm K} $ corresponding to the
28.96\,s period were the magnetospheric radius, the so called `fastness
parameter' (see e.g.  Frank, King \& Raine 1992) would be
\begin{equation}
\omega_{\rm s} \equiv {P_{\rm K} \over P_{\rm s}}= 26.3.
\label{fast1}
\end{equation}
Since  $\omega_{\rm s} \gg 1$, matter transferred from the secondary
could not be accreted onto the white dwarf and would rather be ejected
(as noticed by P98).  WZ Sge could then be similar to another DQ Her
star, AE Aqr which is a discless `ejector' (Wynn, King \& Horne 1997).
Such a solution would be rather attractive: two, of three known, DQ Her
stars would be discless magnetic ejectors; apparent similarities
between \WZ and AE Aqr have already been already pointed out in P80.

We think, however, that such a model cannot describe WZ Sge. First,
Wynn et al. (1997) show that in AE~Aqr the H$_\alpha$ Doppler maps of
the system are consistent with their simulations of a discless flow.
Similar maps for \WZ (Spruit \& Rutten 1998) look completely different
and clearly show the presence of a disc and not just a ring, although
the authors point out that the brightness distribution is less
`disclike' than in other CVs. One should bear in mind, however, that
as pointed out by Spruit \& Rutten \shortcite{sr98} the H$_\alpha$
brightness may not be a good tracer of matter distribution. On the other hand
Mennickent \& Arenas (1998) find that the accretion flow in WZ Sge forms
a `ring' with a ratio of the inner to the outer radii $\sim 0.3$.

Second, contrary to AE Aqr, \WZ is a dwarf nova so that we expect an
accretion disc to be present prior to, and during the outburst (this is
independent of the validity of the DIM - there is ample observational
evidence that dwarf nova outbursts require the presence of accretion
discs around white dwarfs). One could imagine that enhanced mass
transfer prior to outburst could squeeze the magnetosphere and allow
the reappearance of an accretion disc. However, even if this happened,
the superoutbursts observed in WZ Sge would not occur. This is shown in
Figures 2-4 where we have plotted the magnetospheric radius as a
function of the accretion rate for different primary masses.  For the
magnetospheric radius we use the approximate formula (see e.g.  Frank
et al. 1992)
\begin{equation}
R_{M}= 9.8 \times 10^8 \dot M_{15}^{-2/7} M_1^{-1/7} \mu_{30}^{4/7} {\rm cm}
\label{rm}
\end{equation}
where $\dot M_{15}$ is the accretion rate in $10^{15}$ g s$^{-1}$ and 
$\mu = \mu_{30} (10^{30} {\rm gauss\ cm^3}) = B_*R_{1}^3$ is the magnetic 
moment of a white dwarf with a surface field $B_*$. Magnetic moments
corresponding to the 29.69 s oscillation are: $\mu_{30}=34$ for $M_1=0.45$,
$\mu_{30}=55$ for $M_1=0.80$ and $\mu_{30}=77$ for $M_1=1.2$ (see also
Section 3.4).

Near maximum of the 1978/1979 outburst most of the disc material was
accreted onto the white dwarf. According to Smak (1993) the maximum
accretion rate during the 1978 outburst was $3.2 \times 10^{18}$ g
s$^{-1}$. For primary masses larger than Smak's $M_1=0.45$, this
maximum would be lower (Smak - private communication).  As the mass
accretion increases during the rise to outburst the magnetospheric
radius moves inward. Accretion occurs  only when this radius moves
inside the corotation radius. Since no clear oscillations are seen
during most of the outburst (Patterson et al.,\ 1981) this suggests
that in WZ Sge the disc was able to reach the white dwarf surface
(P98).  It can be seen, however, from Figures 2-4 that at outburst
maximum the magnetospheric radius given by Eq. (\ref{mrad1}), i.e.  the
inner disc radius, would just reach the corotation radius.  For a given
maximum accretion rate this is the case for all white dwarf masses
(because the relevant radii are Keplerian) but in reality, for higher
masses, the magnetospheric radius would not even get there, since in
this case the maximum accretion rate would be smaller than the value
plotted on Fig.  2-4. It is obviously impossible to have the maximum of
accretion luminosity at the very moment at which accretion onto the
white dwarf just begins.

We conclude therefore that \WZ is not in a discless ejector phase. 
As we will show below, \WZ is most probably in a state intermediate 
between those of the other two DQ Her stars: AE Aqr - a pure 
discless ejector (Wynn et al., 1997) and DQ Her itself which seems to
have a steady accretion disc \cite{w95}.

\subsection{Blobs at the outer disc edge}

As can been seen in Figure 2, the Keplerian radius corresponding to the
frequency giving the 28.96\,s beat with the white dwarf spin frequency
is very close to the outer disc radius as determined from modelling
observational data. For $M_1=0.45$  it coincides, within the error bars,
with $R_D$ and one can expect that this will be true  also for white dwarf
masses smaller than $\sim 1 M_{\odot}$. We propose, therefore, that the
28.96 s beat frequency results from reprocessing of the spinning,
magnetized, white dwarf beam on plasma blobs orbiting at, or close to,
the outer disc radius. According to Spruit \& Rutten (1998) the `tail'
of the observed hot spot's H$_\alpha$ emission is probably due to
material orbiting at the outer edge of the disc at the Keplerian
velocity. This material had crossed the stream-disc interaction region
undergoing a sequence of heating and cooling events that are likely to
form a non-uniform flow. In a stationary accretion disc the outer
radius is well defined and constant in time so that one can expect the
motion of plasma blobs there to give coherent oscillations. Warner
(1995b) suggested that QPOs observed during some dwarf nova outbursts
could be due to blobs orbiting at the outer disc radius. During
outbursts, however, the outer disc radius moves (in and out; see e.g.
Smak 1984a) and so coherent oscillations should not be expected in such a
case.

Our identification of the 28.96 s period origin is therefore
qualitatively consistent with the Spruit \& Rutten (1998) model of WZ
Sge.  A quantitative agreement will be difficult to achieve because the
interpretation of the H$_\alpha$ `tail' as due to matter in Keplerian
motion leads to their high white dwarf mass(1.2 ${\rm M_{\odot}}$). With
such a mass the Keplerian period of the outer disc rim cannot be $\sim
733$ s.  The value of the white dwarf mass in \WZ is, however, subject
to controversy (see e.g. the discussion in Spruit \& Rutten, 1998) but
for our purpose (see below) the value of 1.2 ${\rm M_{\odot}}$ is too
high. A lower value of $\sim 0.8$ ${\rm M_{\odot}}$ would be consistent
with our hypothesis concerning the origin of the 28.96 s period and the
estimates of P98.

\subsection{A (marginally) stable accretion disc ?}

As shown by Lasota et al. (1995) the very long outburst recurrence time
of \WZ can be reconciled with the `standard' values of the viscosity
parameter $\alpha$ if the disc is truncated so that between outbursts
it is (marginally) stable.  As we will see below,  at present, the flow
of transfered matter in WZ Sge does not, probably,  form a standard
{\sl accretion} disc.  We know however that in the past there was an
accretion disc in WZ Sge (and we expect one to form in the future) so
it is interesting to see what is the inner radius required by the
stability criterion.

The critical value of accretion rate $\dot M$$_{\rm crit}$ below which
a disc is in cold, stable equilibrium can be expressed as (Hameury et
al.,\ 1998; for similar expression see e.g. Ludwig, Meyer-Hofmeister 
\& Ritter 1994; Smak,1984b):
\begin{equation}
\dot{M}_{\rm crit,15} = 4.0\,\, \alpha^{-0.04}\,\, {M}_{1}^{-0.89}
\,\, R_{10}^{2.67},
\label{mdotcr}
\end{equation}
where $\dot M$$_{\rm crit,15}$ is the critical $\dot M$ in units of
10$^{15}$\,g\,s$^{-1}$ and $R_{10}$ the accretion disc radius in units
of 10$^{10}$\,cm. Note that $\dot M$$_{\rm crit}$ depends only weakly
on the value of $\alpha$ (in similar formulae by other authors the
critical $\dot M$ is independent of $\alpha$ [e.g. Ludwig et al., 1994]
- it should be kept in mind that Eq. (\ref{mdotcr}) is a fit to numerical 
results).

It follows from Eq. (\ref{mdotcr}) that for a given $\dot M$, an
accretion disc will be (marginally) stable for $r \gta R_{\rm crit}$
where
\begin{equation}
R_{\rm crit,10} = 0.6\,\, \dot M_{15}^{0.375}\,\, \alpha^{0.015}\,\, 
{M}_{\rm wd}^{0.333}.
\end{equation}
The corresponding Kepler frequency at $R_{\rm crit}$ is given by
\begin{equation}
\omega_{\rm crit} = 0.025\,\, \dot{M}_{15}^{-0.56}\,\, \alpha^{-0.022}.
\end{equation}

This equation shows that $\omega_{\rm crit}$ is independent of the mass
of the white dwarf and only very weakly dependent on $\alpha$. Since the
quiescent $\dot M$ in WZ\,Sge is $\sim 10^{15}$\,g\,s$^{-1}$ (Smak,
1993) this leads to a Kepler period at $R_{\rm crit}$, i.e.\ $P(R_{\rm
crit}) \sim 330-370$ \,s for values of $\alpha \sim $ 0.01--1.
Therefore, in the framework of the truncated disc model, any beat
period between the white dwarf spin period and a Kepler period of
matter in the accretion disc in the quiescent state of WZ\,Sge should
have periods shorter than $\sim$30.4\,s, as observed.

We notice that the recently re-appeared oscillation at 29.69\,s (P98)
is very close to this value. We therefore suggest that the 29.69\,s
period is the beat between the white dwarf spin period and the Kepler
period of matter near $R_{\rm crit}$. Since the inner disc radius,
$R_{\rm in}$ is larger than $R_{\rm crit}$ in quiescence, the 29.69\,s
would mark the inner part of the disc, where matter is dominated by the
magnetic field.

We propose that the magnetospheric radius is located at 
\begin{equation}
R_M(29.69)= 8.87 \times 10^9 M_1^{1/3}\ {\rm cm}
\label{magnr2}
\end{equation}
corresponding to the Keplerian period $P_M= 454.65$ s. The fastness
parameter is now 
\begin{equation}
\omega_{\rm s} \equiv {P_{\rm K} \over P_{\rm s}}= 16.3,
\label{fast2}
\end{equation}
still much larger than 1 so that most of the mass transfered from the
secondary cannot be accreted by the white dwarf, but is ejected from
the system (see below). 

Not all matter is ejected - a few percent finds its way to the white
dwarf and accounts for the pulsed emission.  The configuration we have
in mind is different from disc accretors or disc accretors with
overflowing  accretion stream (see e.g.  Frank, King \& Lasota, 1987;
Hellier, 1996). It is also different from the pure, discless ejector
models of King (1993) and Wynn \& King (1995). It could be a truncated
`excretion disc'.  It might be a structure one obtains in the Wynn \&
King (1995) model for low values of the drag coefficient $k_0$.
According to these authors $k_0 \propto B^2 P_{\rm orb}$ so that $k_0$
for WZ Sge should indeed be lower than in AE Aqr.

From Eqs. (\ref{rm}), (\ref{magnr2}) and the white dwarf mass-radius
relation one can deduce that for higher ($\gta 0.7 M_{\odot}$) white
dwarf masses, at outburst maximum the magnetosphere would not reach
the white dwarf surface.  Even taking into account the rather crude
description of the disc - magnetic field interaction used above, one
can conclude that our model is incompatible with high white dwarf
masses.

\subsection{The magnetic field of WZ Sge}

If the inner disc radius is given by $R_{\rm in}= R_M$ with $R_M$ set
by the observed 29.69\,s oscillation, we derive magnetic field
strengths of $\sim 3.6~\times 10^4$\,G, $\sim 1.6~\times 10^5$\,G,
and $\sim 1.3~\times 10^6$\,G, respectively for $M_1=0.45$, $M_1=0.8$
and $M_1=1.2$. In the last case the white dwarf in \WZ would have a
magnetic field surface intensity comparable to that of most IPs. The
magnetic moment of the white dwarf in \WZ is therefore $\mu \sim 3 - 8
\times 10^{31}$ G cm$^3$.

If the magnetospheric radius is given by Eq. (\ref{magnr2}) then, for all
white dwarf masses of interest,
\begin{equation}
R_M > 4 R_1
\label{wwfcr}
\end{equation}
which, according to Wickramasinghe, Wu \& Ferrario \shortcite{wwf} (see also
Warner, 1995b) means that WZ Sge should be a hard X-ray emitter, as it is
indeed observed. In this respect, as mentioned before, WZ Sge is different from
the other two DQ Her stars, but for a reason that is obvious: a much lower 
mass transfer rate.

On the other hand, in outburst, WZ Sge is not a DQ Her star but enters
the dwarf nova oscillation (DNO) regime in which the `slippage of the
surface', i.e. differential rotation white dwarf's outer layers is
expected  (Warner, 1995a,b). For WZ Sge the minimum field required to
maintain rigid rotation is $B_{\rm min} \sim 2 \times 10^6
M_1^{5/2}$~G, so in WZ Sge $B < B_{\rm min}$ (Figure 5). No confirmed
DNOs have been seen, however, during the last outburst \cite{p81}.
\begin{figure} 
\centering\epsfig{figure=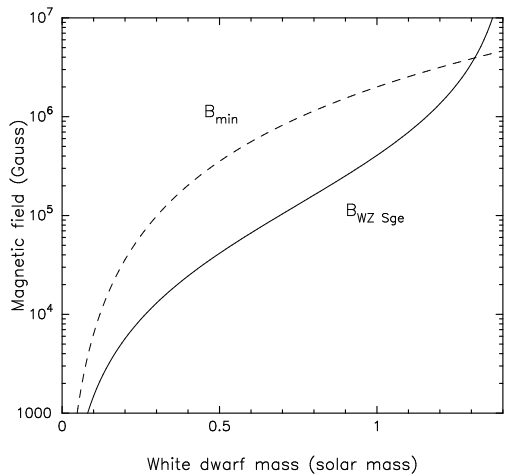,width=7.5cm,angle=0} 
\caption{The magnetic field (continuous line) of WZ Sge calculate assuming
that the magnetospheric radius is given by Eq. (\ref{magnr2}). 
Nauenberg's mass-radius relation is also assumed. The dashed line
represent the minimum magnetic field $B_{\rm min}$ required to maintain
rigid rotation of the white-dwarf outer layers during outburst.
}
\end{figure}

\section{The accretion history of \WZ}

If \WZ is in an ejector phase, its white dwarf should be spinning down.
Indeed, Patterson (1980) reports  $\dot P_s = 8 \times 10^{-12}$.  This
value is very large compared with the $\dot P_s = 5.64 \times 10^{-14}$
observed in AE Aqr (de Jager et al., 1994). The implied spindown power
of \WZ would be $ \gta 5 \times 10^{35} {\rm erg\ s^{-1}}$, two orders
of magnitude higher than in AE Aqr.  Wynn et al. (1997) find that
(assuming that plasma interacts with magnetic field in the form of blobs)
$L_{\rm spin}$ scales with $\dot M$. Since in \WZ the mass transfer rate is
two orders of magnitude {\sl lower} than in AE Aqr the two orders of
magnitude higher spindown power is hard to understand.  One
should note, however, an interesting point: in \WZ the light cylinder
radius $R_c\equiv c/\omega = 1.3 \times 10^{11} {\rm cm}$ is larger
than the size of the system ($a \sim 3.4 \times 10^{10} {\rm cm}$)
which allows, in principle, acceleration of particles outside the
system.  It may well be that results for systems in which $R_c \lta a$
do not apply to WZ Sge. For particles accelerated to speeds close to
the speed of light
\begin{equation}
L_{\rm spin,c}\approx \dot M R^2_c \omega^2 \approx 9 \times 10^{35}
\dot M_{15}\ {\rm erg\ s^{-1}}
\end{equation}
so that to obtain the spindown power that corresponds to $\dot P \sim 8  
\times 10^{-12}$ the escape velocity of particles removing rotational
energy would have to be a significant ($\sim 90\%$) fraction of the
speed of light. This is, probably, not totally absurd since TeV
$\gamma$--rays have been observed from AE Aqr (but not from WZ Sge which
is closer) but we would still expect a spin-down rate in WZ Sge closer to
(or lower) than that measured in AE Aqr. A reliable $\dot P$ would be
an important test of any model assuming that WZ Sge contains a fast
magnetic rotator.

In any case we have to consider how the white dwarf in \WZ had been
spun up to its present rate. The only way to achieve this is 
through accretion disc spin-up. The white dwarf spin rate increases
as it accretes the Keplerian angular momentum of matter at the
inner disc edge. An equilibrium period is reached when  angular 
momentum is accreted at the same rate as it is centrifugally expelled
by the spinning white dwarf (a magnetic field is not necessary for
this to work). In the case of a magnetized white dwarf the equilibrium
period is equal to the Keplerian period of the magnetosphere 
($\omega_s=1$). The equilibrium spin is then 
\begin{equation}
P_{\rm eq} = 360 M_1^{-5/7} \dot M_{15}^{-3/7} \mu_{31}^{6/7} \ {\rm s}
\label{peq}
\end{equation}

We see immediately that an accretion rate $\sim 10^{17}$ g s$^{-1}$ is
required to bring \WZ to its present rapid rotation rate. Such
accretion rates are typical of nova-like CVs at orbital periods longer
than 3 hours whereas \WZ is very close to the minimum period for CVs
where secular mass-transfer rates are two orders of magnitude lower.
The present mass transfer rate in WZ Sge is in good agreement with that
predicted by mass loss from the secondary being driven by gravitational
radiation alone. It might even be possible that WZ Sge is already on
the `other side' of the minimum period (see e.g. Patterson, 1998). At
orbital periods close to 80 minutes the secondary, which due to
mass-loss is out of thermal equilibrium, stops to contract in response
to mass loss and the binary system starts expanding to longer orbital
periods. This `bounce' is  helped by the secondary becoming degenerate
(see e.g. King, 1988). It would seem, therefore, that high accretion
rates are impossible at short orbital periods.

The situation, however, is not as hopeless as it would seem. We are
interested here in time-scales of the order of the white dwarf spin-up
time, i.e. 10$^4-10^5$ years. Fluctuations on such short time-scale do
not modify the secular evolution of the binary.

There is at least one  piece of evidence which shows that the accretion
history of \WZ might have had episodes of high mass-transfer.  There
exist, at orbital periods between 79 and 92 minutes systems which are
supposed to have high ($\sim 10^{17}$ g s$^{-1}$) mass transfer rates.
These are ER UMa systems which show extremely short intervals between
superoutbursts (19 -- 44 d) and very short (3 - 4 d) `normal' outburst
interval.  They seem to be the high accretion rate equivalent of the
low accretion rate systems showing superoutbursts: the SU UMa's (WZ Sge
is a SU UMa system showing superoutbursts only).  The idea that DI UMa
(one of the 4 known ER UMa's with $P_{\rm orb} =79$ min), which is more
luminous than WZ Sge by a factor $\sim 50$, spends most of its life as
`an ordinary WZ Sge star', but was caught during an `upward surge in
accretion', has been proposed by Patterson (1998).  He speculates that
the short period CV mass-transfer cycles in which $\dot M$ surges to
high values ($\sim 1 - 5 \%$ of the cycle time) are somehow associated
with classical nova eruptions because in addition to the 4 ER UMa's, 4
other systems are `too bright' for their orbital periods and 3 of them
are classical nova remnants.

Whatever the reason, what is important for our argument is the presence
of systems which accrete at high rates but have orbital periods close
to that of WZ Sge. With such accretion rates the white dwarf in \WZ can be
spun up to $P_{\rm eq} = 27.87$ s.

When the white dwarf gets to this equilibrium spin rate the mass
transfer is stable (Wynn \& King, 1995). After some time the mass
transfer rate will return to its secular value.  This value, however,
is two orders of magnitude lower than that required for $P_{\rm eq}\sim
{\rm few} \times 10$ s.  In a very short time (viscous time of the disc
$\sim$days) the magnetosphere will start expanding and becoming larger
than the corotation radius. The system enters into the ejector phase.
As shown by Wynn \& King (1995) this phase is dynamically stable (for
$0.05 \lta q \lta 1$). It will stay in this, WZ Sge-phase (for about
$10^5$ years) until it gets to a new spin equilibrium corresponding to
the low accretion rate ($\sim$ few minutes). Such a system has been
recently observed:  IP RX J0757.0+6306 \cite{tovm} has an orbital
period very close to that of WZ Sge ($81\pm 5 {\rm min}$) but its spin
period $8.52\pm 0.15$ min corresponds to the equilibrium value as given
by Eq (\ref{peq}) for the expected secular mass-transfer rate.  The
white dwarf's magnetic field in this system could have, therefore, a
strength close to that of WZ Sge. The binary RX J0757.0+6306,
therefore, could be a WZ Sge-type system at a particular phase of its
spin-up/spin-down history. (Note that this system is different from the
other IP with a very short orbital period (84 min), RX J1238-38.  There
the spin period is 36 min \cite{bcrw} so it would be a `classical' IP
with $\mu_{31} \sim 10$.)

A jump in the mass transfer rate could bring such an equilibrium system
back to a spin-up phase. If RX J0757.0+6306 has indeed a magnetic field
close to the one we obtained for WZ Sge and if its mass-transfer rate
is close the secular one, its accretion disc should be truncated at a
radius close to the value given by Eq. (\ref{magnr2}) and therefore
stable with respect to the thermal-viscous instability. It is not yet
clear if this system shows any kind of dwarf-nova activity.

\section{Summary of the model}

We can summarize our model (or `scenario') as follows: WZ Sge
contains a $\lta 0.8 M_{\odot}$ magnetized white dwarf spinning with a
period of 27.87 s. The magnetic moment is $\lta 5 \times 10^{31}$ G
cm$^3$, or magnetic field at the surface $\lta 2 \times 10^5$ G.

In quiescence the mass transferred from the secondary forms a
disc--like structure with an outer radius and inner radii at 1.22
and $8.87 \times 10^{9} M_1^{1/3}$ cm
respectively. The accretion flow is disrupted by the rapidly rotating
magnetosphere and since the fastness parameter is $\gg$ 1 (the
corotation radius is much smaller than the magnetospheric radius) most
of the matter cannot be accreted by the white dwarf. A small fraction
($\lta 10\%$) finds its way to the white dwarf surface where it is
responsible for the observed X--ray emission.

The white dwarf is in a spin-down phase and most of the transferred
matter is ejected from the system. The disc-like structure formed by the
flow might be composed of plasma `blobs'. At outburst, which is triggered
by an enhanced mass transfer event, an accretion disc is (re)created. At
outburst maximum the disc reaches down to the white dwarf's surface.

The white dwarf in WZ Sge will spin down to an equilibrium  rotation
period of a few minutes. In this phase an increase of accretion rate
will transform it into an ER UMa star.  After a spin-up phase the
system will become a DQ Her star once more.


\subsection*{Acknowledgments} We thank Jean-Marie Hameury, Henk Spruit
and Joe Smak for helpful discussions and help.  Part of this work was
performed during a visit of JPL to Oxford and EK to Meudon. They thank
the respective departments for their hospitality.  The visit of JPL to
Oxford and EK to Meudon was made possible by the British-French
collaboration grant {\sl Alliance}.

\label{lastpage}


\begin{thebibliography}{}

\bibitem[\protect\citename{Buckley et al., }1998]{bcrw}Buckley, D.A.H., 
        Cropper, M., Ramsay, G., Wickramasinghe, D.T. 1998, MNRAS, 299, 83
\bibitem[\protect\citename{Cannizzo, }1993]{can93}Cannizzo J.K., 1993, in
	Accretion in Compact Stellar Systems, ed. J. Wheeler,
	World Scientific, p. 6
\bibitem[\protect\citename{Cheng et al., }1997]{ch97}Cheng, F.H., Sion, 
        E.M., Szkody, P., Huang, M. 1997, ApJ, 484, L149
\bibitem[\protect\citename{de Jager et al., }1994]{dJ}de Jager et al., 1994,
        MNRAS, 267, 577
\bibitem[\protect\citename{Frank et al., }1987]{fkl} Frank J., King A.R.,
        Lasota J.-P. 1987, A\&A, 178, 137
\bibitem[\protect\citename{Frank et al., }1992]{APiA} Frank J., 
	King A.R., Raine D., 1992, Accretion Power in Astrophysics. 
	CUP 
\bibitem[\protect\citename{Gammie \& Menou, }1998]{gm98} Gammie C.F., Menou K.,
	1998, ApJ, 492, L75
\bibitem[\protect\citename{Hameury, Lasota \& Hur\'e, }1997]{hlh97} Hameury
	J.-M., Lasota J.-P., Hur\'e J.-M., 1997a, MNRAS, 287, 937
\bibitem[\protect\citename{Hameury et al., }1998] 
	{hmdl98} Hameury J.-M., Menou K., Dubus G., Lasota J.-P., Hur\'e J.-M., 
        1998, MNRAS, 298, 1048 
\bibitem[\protect\citename{Hellier, }1996]{h96} Hellier, C. 1996, COSPAR, Adv.
        Sp. Res. in press
\bibitem[\protect\citename{Hessman et al.,} 1984]{h84}
        Hessman F.V., Robinson E.L., Nather, R.E., Zhang, E.-H., 1984, 
        ApJ, 286, 747
\bibitem[\protect\citename{King, }1988]{k88}King, A.R. 1988, QJRAS, 29, 1
\bibitem[\protect\citename{King, }1993]{k93}King, A.R. 1993, MNRAS, 261, 144
\bibitem[\protect\citename{King, }1997]{k97}King, A.R. 1997, in Relativistic 
        Gravitation and Gravitational Radiation, eds. J.-A. Marck \& J.-P.
        Lasota, CUP, p. 105 
\bibitem[\protect\citename{Lasota \& Hameury, }1998]{lh98}
	Lasota J.-P., Hameury J.-M., 1998, in Accretion Processes in
	Astrophysical Systems - Some Like it Hot, Proceedings of the 8th
	Annual October Conference in Maryland, eds. S.S. Holt \& T.R
        Kallman, AIP, p. 351
\bibitem[\protect\citename{Lasota et al., }1995]{lhh95} Lasota J.-P.,
	Hameury J.-M., Hur\'e, J.-M., 1995, A\&A, 302, L29
\bibitem[\protect\citename{Livio \& Spruit, }1991]{ls91}
        Livio, M., Spruit, H.C. 1991, A\&A, 252, 189
\bibitem[\protect\citename{Ludwig et al., }1994]{lmhr}
        Ludwig, K., Meyer--Hofmeister, E. \& Ritter, H. 1994,  A\&A, 290, 473
\bibitem[\protect\citename{Mennickent \& Arenas}1998]{ma98}Mennickent, R.E. \& 
        Arenas, J. 1998, PASJ, 50, 333 
\bibitem[\protect\citename{Meyer \& Meyer-Hofmeister, }1994]{mm94} Meyer F.,
	Meyer-Hofmeister E., 1994, A\&A, 288, 175
\bibitem[\protect\citename{Meyer-Hofmeister et al., }1998]{mhml} 
        Meyer-Hofmeister E., Meyer, F., Liu, B.F. 1998, A\&A, in press
\bibitem[\protect\citename{Mineshige et al., }1998]{mlmmh}Mineshige, S., Liu,
        B.F., Meyer, F., Meyer-Hofmeister E. 1998, PASJ, 50,
\bibitem[\protect\citename{Nauenberg, }1972]{n72}Nauenberg, M. 1972,
         ApJ, 175, 417
\bibitem[\protect\citename{Osaki, }1996]{o96}Osaki, Y. 1996, in Cataclysmic 
        Variables and Related Objects; IAU Coll.
        158, eds. A. Evans \& J.H. Wood, p. 127
\bibitem[\protect\citename{Patterson, }1980]{P80}Patterson, J. 1980,
        ApJ, 241, 235 (P80)
\bibitem[\protect\citename{Patterson, }1998]{jp98}Patterson, J. 1998,
        PASP, 110, 1132
\bibitem[\protect\citename{Patterson et al., }1981]{p81}Patterson J., McGraw,
        J.T., Coleman, L., Africano, J.L. 1981, ApJ, 248, 1067
\bibitem[\protect\citename{ferson et al., }1998]{prkm}Patterson J.,
	Richman, H., Kemp J., Mukai K. 1998, PASP, 110, 403 (P98)
\bibitem[\protect\citename{Robinson et al., }1978]{rnp}Robinson, E.R.,
        Nather, R.E., Patterson J. 1978, ApJ. 219, 168
\bibitem[\protect\citename{Shakura \& Sunyaev }1973]{ssu} Shakura N.I., 
	Sunyaev R.A., 1973, A\&A, 24, 337
\bibitem[\protect\citename{Smak, }1984a]{sm84a} Smak, J.I. 1984a, Acta Astron., 34,
        93 
\bibitem[\protect\citename{Smak, }1984b]{sm84b} Smak, J.I. 1984b, Acta Astron., 34,
        161
\bibitem[\protect\citename{Smak, }1993]{sm93} Smak, J.I. 1993, Acta Astron., 
        43, 101
\bibitem[\protect\citename{Smak, }1997]{sm97} Smak, J.I. 1997, in Cataclysmic 
        Variables and Related Objects; IAU Coll.
        158, eds. by A. Evans \& J.H. Wood, p. 45
\bibitem[\protect\citename{Spruit \& Rutten, }1998]{sr98}Spruit, H.C., Rutten,
        R. G. M. 1998, MNRAS 299, 768
\bibitem[\protect\citename{Tovmassian et al., }1998]{tovm}Tovmassian, G.H., 
        et al. 1998, A\&A, 335, 227 
\bibitem[\protect\citename{Warner, }1986]{w86} Warner B., 1986, MNRAS, 219, 347
\bibitem[\protect\citename{Warner, }1995a]{w95} Warner B., 1995a,
	Cataclysmic Variable Stars, CUP, Cambridge
\bibitem[\protect\citename{Warner, }1995b]{w95b} Warner B., 1995b,
	in Cape Workshop on Cataclysmic Variables; eds. D.A.H. Buckley \& 
        B. Warner, ASP Conf. Series, Vol. 85, p. 343
\bibitem[\protect\citename{Warner et al., }1996]{wlt96} Warner B., Livio, M.,
        Tout, C.A. 1996, MNRAS, 282, 735
\bibitem[\protect\citename{Welsh et al., }1997]{wlsh97}Welsh, W.F.,
         Skidmore, W., Wood, J.H., Cheng, F.H., Sion, E.M. 1997, MNRAS, L57
\bibitem[\protect\citename{Wickramasinghe et al., }1991]{wwf}
        Wickramasinghe, D.T., Wu, K., Ferrario, L. 1991, MNRAS, 249, 460
\bibitem[\protect\citename{Wood, }1999]{jw99} Wood, J.H. 1999, in Disk Instabilities 
        in Close Binary Systems - 25 years of the Disk Instability Model, 
        eds. S. Mineshige S. \& J.C. Wheeler, Universal Academy Press, in press
\bibitem[\protect\citename{Wynn \& King, }1992]{wk92}Wynn, G.A., King,
        A.R. 1992, MNRAS, 255, 83
\bibitem[\protect\citename{Wynn \& King, }1995]{wk95}Wynn, G.A., King,
        A.R. 1995, MNRAS, 275, 9
\bibitem[\protect\citename{Wynn et al., }1997]{wkh97}Wynn, G.A., King,
        A.R., Horne, K. 1997, MNRAS, 286, 436
\end{thebibliography}
\end{document}